\documentclass[12pt]{article}
\pdfoutput=1
\usepackage{graphicx}
\catcode`\@=11

\global\arraycolsep=2pt
\usepackage{amsbsy,amssymb,latexsym,amsfonts,amsmath}
\usepackage{graphicx,color}

\begin{document}
\begin{flushright}
{RUP-16-13, IPMU16-0069}
\end{flushright}

\vspace*{0.7cm}

\begin{center}
{ \Large Bootstrap bound for conformal multi-flavor QCD on lattice \\}
\end{center}
\vspace*{1.0cm}
\begin{center}
{Yu Nakayama}
\vspace*{1.0cm}

Department of Physics, Rikkyo University, Toshima, Tokyo 171-8501, Japan

and 

Kavli Institute for the Physics and Mathematics of the Universe (WPI),  
\\ University of Tokyo, 5-1-5 Kashiwanoha, Kashiwa, Chiba 277-8583, Japan

\vspace{3.8cm}
\end{center}

\begin{abstract}
The recent work by Iha et al shows an upper bound on mass anomalous dimension $\gamma_m$ of  multi-flavor massless QCD at the renormalization group fixed point  from the conformal bootstrap in $SU(N_F)_V$ symmetric conformal field theories under the assumption that the fixed point is realizable with the lattice regularization based on staggered fermions.  We show that the almost identical but slightly stronger bound applies to the regularization based on Wilson fermions (or domain wall fermions) by studying the conformal bootstrap in $SU(N_f)_L \times SU(N_f)_R$ symmetric conformal field theories. For $N_f=8$, our bound implies $\gamma_m < 1.31$ to avoid dangerously irrelevant operators that are not compatible with the lattice symmetry. 


\end{abstract}

\thispagestyle{empty} 

\setcounter{page}{0}

\newpage
\section{Introduction}
Realizing strongly coupled conformal gauge theories on a lattice has been one of the active research arenas in theoretical physics. It would give non-perturbative information on various critical exponents such as mass anomalous dimension $\gamma_m = 3 - \Delta_{\bar{\psi}\psi}$. Finding a model with larger mass anomalous dimension is important for its possible application to particle physics (see e.g. \cite{Sannino:2009za} and reference therein). We note that in four dimensions, as long as we work in Lagrangian quantum field theories, it is necessary to consider gauge  theories to realize any non-trivial renormalization group fixed point, and the Monte-Carlo simulation to probe such non-trivial renormalization group fixed points is thus much more difficult than in the lower dimensions.

One of the simplest examples of possible conformal gauge theories in four dimensions is multi-flavor massless QCD \cite{Caswell:1974gg}. While we have a better theoretical control in the weakly coupled regime \cite{Banks:1981nn} (so called Banks-Zaks fixed point in the literature), we have two unsolved issues in the strongly coupled regime: (1) what is the critical number of flavors with a given gauge group? (2) what is the maximal value of the mass anomalous dimension?

We concentrate on the second issue here. There is an absolute bound on the mass anomalous dimension  $\gamma_m \le 2$ from the unitarity bound of conformal field theories \cite{Mack:1975je}.\footnote{Throughout the work, we assume that the renormalization group fixed point shows conformal invariance rather than mere scale invariance, which is true in the weakly coupled regime \cite{Polchinski:1987dy}. See e.g. \cite{Nakayama:2013is} for the validity of this assumption.} There is a proposed more stringent bound $\gamma_m \le 1$ from the analysis of the Schwinger-Dyson equation \cite{Appelquist:1988yc}. Intuitively, the latter corresponds to the bound on the coupling constant where the $SU(N_f)_L \times SU(N_f)_R$ symmetric four-Fermi terms become relevant, inducing the Nambu-Jona-Lasinio mechanism of chiral condensation. Note, however, that unless in the large $N_f$ limit, in which the mass anomalous dimension and the relevance of the four-Fermi interaction may be related, there is no simple relation such as $\Delta_{\mathrm{four-fermi}} = 2\Delta_{\bar{\psi}\psi}$, relating the mass anomalous dimension and the instability due to the four-Fermi terms. Indeed, the approximation used in the Schwinger-Dyson equation may be only justifiable in the large $N_f$ limit. Whether $\gamma_m=1$ or $\gamma_m=2$ is when the conformality will be lost in multi-flavor massless QCD is still under debate \cite{Ryttov:2007cx}\cite{Kaplan:2009kr}.

Along the line of this reasoning, the conformal bootstrap gives more quantitative bound on the mass anomalous dimension in relation to the (ir)relevance of the  $SU(N_f)_L \times SU(N_f)_R$ symmetric four-Fermi terms. While being rigorous, the bound is weak as we will see in the next section and does not exclude the possibility of having larger mass anomalous dimension, in particular in the large $N_f$ limit.

On the other hand, in the recent paper \cite{Iha:2016ppj}, Iha et al addressed a similar but slightly different question: what is the largest mass anomalous dimension of conformal multi-flavor QCD under the assumption that it is realizable with the regularization based on staggered fermions.
The action for the staggered fermion has a smaller symmetry of $U(N_f/4)_L \times U(N_f/4)_R$ than massless QCD in the continuum limit, and we obtain a more constraint from the condition that there is no relevant deformations that cannot be forbidden by the lattice symmetry. It should be noted that their results do not exclude the possibility for multi-flavor massless QCD to realize a fixed point violating their bound. Rather, it means that such a fixed point cannot be approached with the lattice action based on the staggered fermions without further fine-tuning.

In this paper, we address the similar question for the other regularization based on Wilson fermions (or domain wall fermions), in which one can only preserve the $SU(N_f)_V$ subgroup of $SU(N_f)_L \times SU(N_f)_R$ symmetry of the conformal multi-flavor QCD. From the bootstrap analysis in $SU(N_f)_L \times SU(N_f)_R$ symmetric conformal field theories, we will see that the bound is almost identical but slightly stronger.
 While our results, like theirs, do not exclude the possibility of realizing larger mass anomalous dimension in conformal multi-flavor QCD (e.g. for $N_f = 8)$, it implies that if there is any possibility to realize it, we need to tune effective four-Fermi interactions very carefully beyond what we normally do in lattice QCD with the regularization that only preserves $SU(N_f)_V$ symmetry.

\section{Bootstrap program and results}

Our numerical conformal bootstrap program is basically same as the seminal studies in \cite{Rattazzi:2008pe}\cite{Rattazzi:2010yc}\cite{Vichi:2011ux}\cite{Poland:2011ey}\cite{Caracciolo:2014cxa} (but with more complicated symmetry group of $SU(N_f)_L \times SU(N_f)_R$ than $SO(N)$ or $SU(N)$) except that we use the refined implementation of the numerical semidefinite programming  by SDPB \cite{Simmons-Duffin:2015qma}. We consider the four-point functions  $\langle \Phi_{i_Li_R} \bar{\Phi}_{j_Lj_R} \Phi_{k_Lk_R} \bar{\Phi}_{l_Ll_L} \rangle$ of a scalar operator $\Phi_{i_Li_R}$ in the fundamental $\times$ fundamental representation (a.k.a. bifundamental) of $SU(N_f)_L \times SU(N_f)_R$ (i.e. meson operator in QCD) with the scaling dimension $\Delta_{\Phi}$ and their complex conjugates $\bar{\Phi}_{i_Li_R}$. In the following equations, we sometimes use the abbreviation of symmetric traceless $T$, anti-symmetric $A$, and singlet $S$ representation in  each $SU(N_f)$.

The crossing symmetry of the four-point functions gives the following OPE sum rule\footnote{This sum rule was originally obtained for the purpose of bootstrapping the finite temperature QCD. The computation is straightforward but tedious group theoretic combinatorics, which we have implemented with Mathematica based on the general algorithm presented in \cite{Rattazzi:2010yc}.  
	Eventually, we have shifted our focus on the case with $N_f=2$ and $O(n) \times O(m)$ models in \cite{Nakayama:2014lva}\cite{Nakayama:2014sba}. The author would like to thank T.~Ohtsuki for the collaboration.}:
\begin{align}
0 =& \sum_{O\in \Phi \times \bar{\Phi}} \lambda^2_O V_{AdjAdj}^{(+)} + \sum_{O\in \Phi \times {\Phi}} \lambda^2_O V_{TA}^{(-)} + \sum_{O\in \Phi \times \bar{\Phi}} \lambda^2_O V_{AdjAdj}^{(-)} + \sum_{O\in \Phi \times \bar{\Phi}} \lambda^2_O V_{AdjS}^{(+)} + \sum_{O\in \Phi \times {\Phi}} \lambda^2_O V_{AA}^{(+)} \cr
 &+ \sum_{O\in \Phi \times \bar{\Phi}} \lambda^2_O V_{SS}^{(+)} + \sum_{O\in \Phi \times {\Phi}} \lambda^2_O V_{TT}^{(+)} + \sum_{O\in \Phi \times \bar{\Phi}} \lambda^2_O V_{AdjS}^{(-)} + \sum_{O\in \Phi \times \bar{\Phi}} \lambda^2_O V_{SS}^{(-)}
\end{align}
where $(\pm)$ denotes the even $(+)$ or odd $(-)$ spin contributions.
By using the convention
\begin{align}
F & = v^{\Delta_{\Phi}} g_{\Delta_O,l}(u,v) - u^{\Delta_{\Phi}} g_{\Delta_O,l}(v,u) \cr
H & = v^{\Delta_{\Phi}} g_{\Delta_O,l}(u,v) + u^{\Delta_{\Phi}} g_{\Delta_O,l}(v,u)
\end{align}
with the conformal block $g_{\Delta_O,l}$ being normalized as in \cite{Hogervorst:2013kva}, whose explicit expression can be found in \cite{Dolan:2003hv},
the each representation contributes to the sum rule as
\begin{align}
V_{AdjAdj}^{(+)} &= \left( \begin{array}{cc} (1+N_f^{-2}) F \\ (-1+N_f^{-2}) H \\  -2N_f^{-1} F  \\ F \\ -H \\ N_f^{-2} F \\ N_f^{-2} H \\ -N_f^{-1}F \\ - N_f^{-1} H  \\ 
\end{array} \right) \ , \ \ 
V_{TA}^{(-)} = \left( \begin{array}{cc} 0 \\ 0 \\  0  \\ -F \\ -H \\ F \\ - H \\ 0 \\ 0  \\ 
\end{array} \right) \ , \ \ 
V_{AdjAdj}^{(-)} = \left( \begin{array}{cc} (1+N_f^{-2}) F \\ (-1+N_f^{-2}) H \\  -2N_f^{-1} F  \\ -F \\ H \\ -N_f^{-2} F \\ -N_f^{-2} H \\ N_f^{-1}F \\ N_f^{-1} H  \\ 
\end{array} \right) \cr
V_{AdjS}^{(+)} &= \left( \begin{array}{cc} -2N_f^{-1} F \\ -2N_f^{-1} H \\ 2 F  \\ 0 \\ 0 \\ -2N_f^{-1}F \\ -2N_f^{-1} H \\ F \\ H  \\ 
\end{array} \right) \ , \ \ 
V_{AA}^{(+)} = \left( \begin{array}{cc} 0 \\ 0 \\  0  \\ F \\ H \\ F \\ - H \\ -F \\ H \\ 
\end{array} \right) \ , \ \
V_{SS}^{(+)} = \left( \begin{array}{cc}  F \\ H \\ 0 \\ 0 \\ 0 \\  F \\ H \\ 0 \\ 0  \\ 
\end{array} \right) \cr
V_{TT}^{(+)} &= \left( \begin{array}{cc} 0 \\ 0 \\ 0  \\ F \\ H \\ F \\ - H \\ F \\ -H  \\ 
\end{array} \right) \ , \ \ 
V_{AdjS}^{(-)} = \left( \begin{array}{cc} -2N_f^{-1}F \\ -2N_f^{-1} H \\  2F  \\ 0 \\ 0 \\ 2N_f^{-1}F \\ 2N_f^{-1}H \\ -F \\ -H \\ 
\end{array} \right) \ , \ \
V_{SS}^{(-)} = \left( \begin{array}{cc}  F \\ H \\ 0 \\ 0 \\ 0 \\  -F \\ -H \\ 0 \\ 0  \\ 
\end{array} \right) 
\end{align}
Here, we have symmetrized the two $SU(N_f)$s because in the sum rule, one cannot distinguish the $SU(N_f)_L$ and $SU(N_f)_R$. We also note that for this four-point function, the sum rule is the same with or without extra $U(1)_A$. In other words, the $U(1)_A$ anomaly cannot be seen in the bootstrap bound we study.\footnote{A possible way to avoid $U(1)_A$ symmetry is to introduce a gap assumption in spin one sector in the singlet representation, but we a priori do not know how much gap should be introduced.}

Our main focus in this paper is $N_f=8$ which is under debate whether the theory possesses a fixed point for the $SU(3)$ gauge group, but we may study the other flavor groups with no difficulty. In this section, we will set $N_f=8$ in the following, but we will comment on the case $N_f=16$ in section 3. 
 A reader who is interested in the other $N_f$ may contact the author. 

The actual implementation of the numerical conformal bootstrap is based on cboot \cite{cboot}, which uses the SDPB as a core part of the semi-definite programming. For the truncation of the search space by number of derivatives, our numerical results in this section are based on $\Lambda (= N_{\mathrm{max}}) = 17$, which is slightly better than the one used in \cite{Iha:2016ppj}. The other parameters such as the number of included spin are chosen appropriately so that the numerical  optimization is stable.

Our first result is the bound on the scaling dimensions of the singlet operator in $SU(N_f)_L \times SU(N_f)_R$ symmetric conformal field theories that appears in the OPE of the scalar operators $\Phi$ and $\bar{\Phi}$. When the scaling dimension of this operator becomes less than four, it is regarded as a dangerously irrelevant operator (because in our explicit models in mind, there is no such relevant operators in the UV continuum limit), and we have to tune the corresponding coupling to realize the fixed point in any regularization in use. We again emphasize  that this does {\it not} mean that such a fixed point cannot exist, but it only means that the realization on the lattice requires further fine-tuning due to the dangerously irrelevant operator.

\begin{figure}[htbp]
	\begin{center}
  \includegraphics[width=12.0cm,clip]{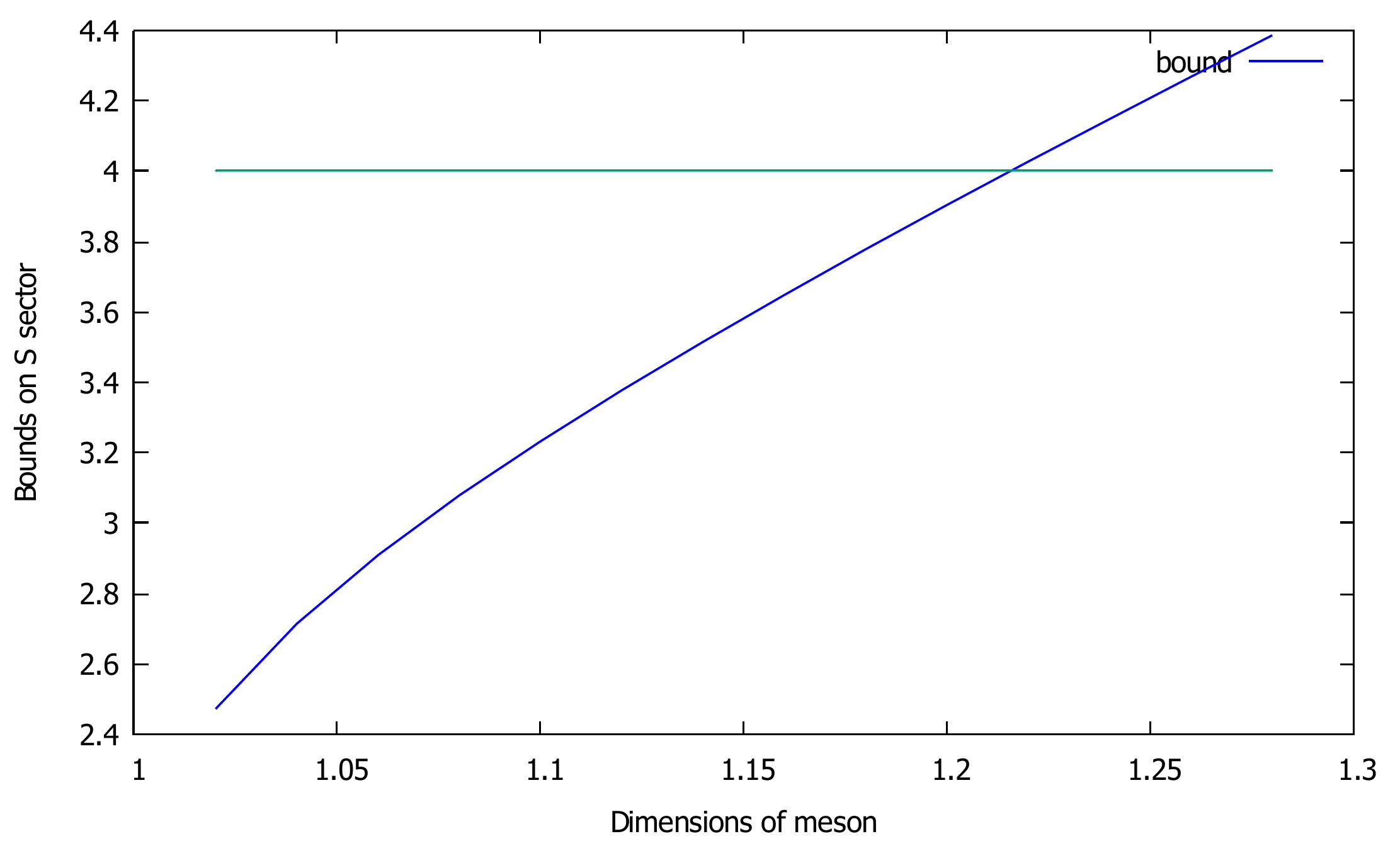}
  \end{center}
  \caption{Bounds on the scaling dimension of operators in the singlet representation.}
  \label{fig:s}
\end{figure}

A plot in Fig \ref{fig:s} shows the bound on the scaling dimensions of the singlet operator for $N_f=8$. This plot directly tells that the bound is $\Delta_{\Phi} > 1.21 $ or $\gamma_m < 1.79 $ in order to avoid the existence of a dangerously irrelevant operator that is singlet. As a side remark, we have found that the bound actually coincides with the bound on the scaling dimensions of the singlet operator in the four-point function of scalar operators in the fundamental (vector) representation  in $SO(128)$ symmetric conformal field theories. This symmetry enhancement is similar to the observation made in \cite{Poland:2011ey}\cite{Nakayama:2014lva}.
 
Let us compare our bound with the one that has been (implicitly) studied in  \cite{Iha:2016ppj}. They studied the bound on the scaling dimensions of operators that appear in the OPE of the scalar operators in the  adjoint  representation of the $SU(N_f)_V$ symmetry. 
Due to the above mentioned symmetry enhancement, the bound on scaling dimension of the singlet operator which they could have computed must be identical to the bound on the scaling dimensions of the singlet operator appearing in the OPE of the scalar operators in the fundamental (vector) representation in $SO(63)$ symmetric CFTs. We have explicitly checked this numerically with the same search space dimension (i.e. same $\Lambda$), which directly shows that our bound is slightly weaker than theirs.

Now, we are going to study the bound on the scaling dimensions of operators in the symmetric traceless $\times$ symmetric traceless  representation and anti-symmetric  $\times$ anti-symmetric  representation because these include singlet scalar operators in the $SU(N_f)_V$, so one may not be able to exclude the corresponding deformations from the effective action without fine-tuning if we use the regularization that only preserves the $SU(N_f)_V$ symmetry (such as Wilson fermions or domain wall fermions). Therefore these operators become dangerously irrelevant if the scaling dimensions become less than four.

\begin{figure}[htbp]
	\begin{center}
  \includegraphics[width=12.0cm,clip]{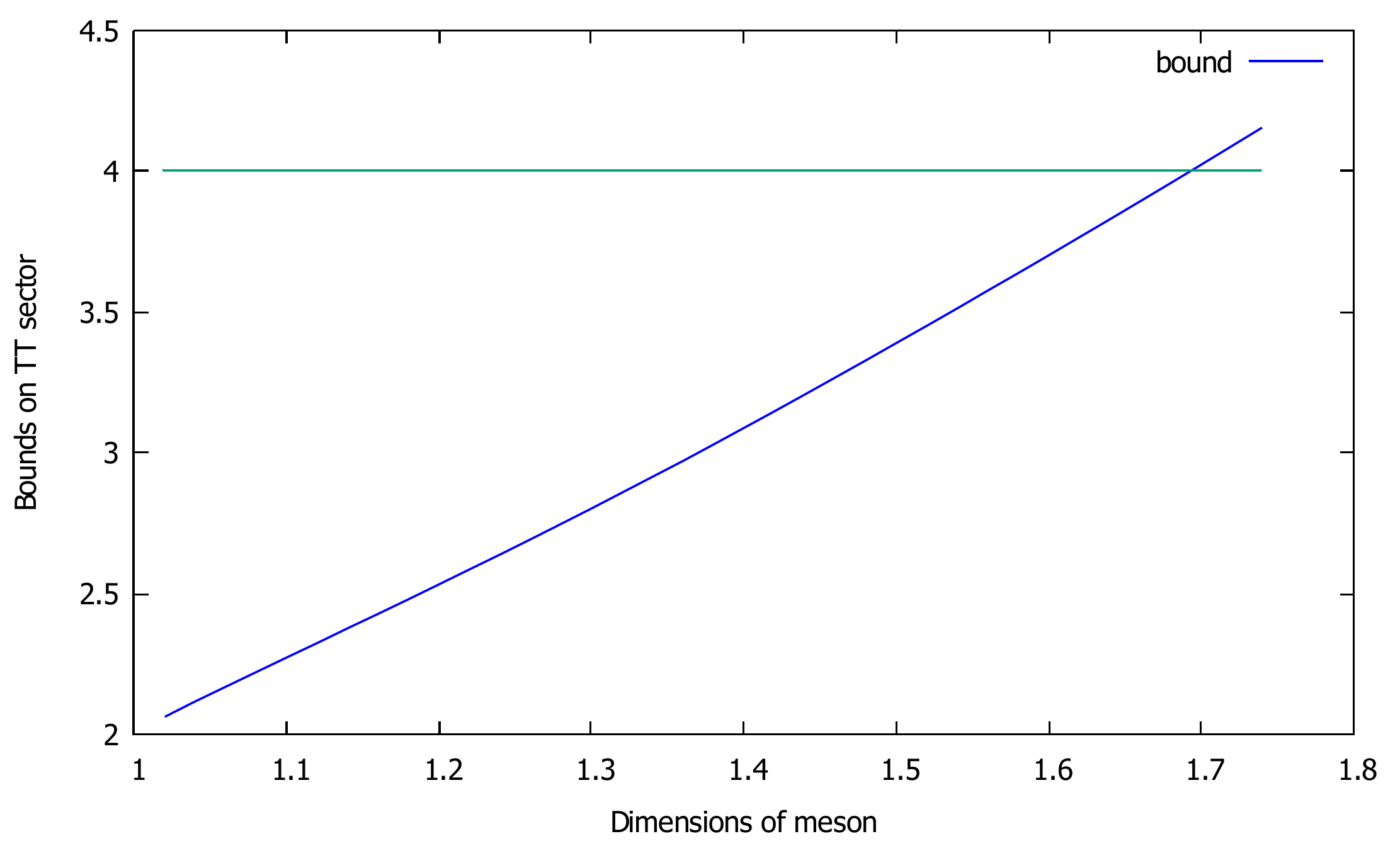}
  \end{center}
  \caption{Bounds on the scaling dimension of operators in the symmetric traceless $\times$ symmetric traceless representation.}
  \label{fig:tt}
\end{figure}

\begin{figure}[htbp]
	\begin{center}
  \includegraphics[width=12.0cm,clip]{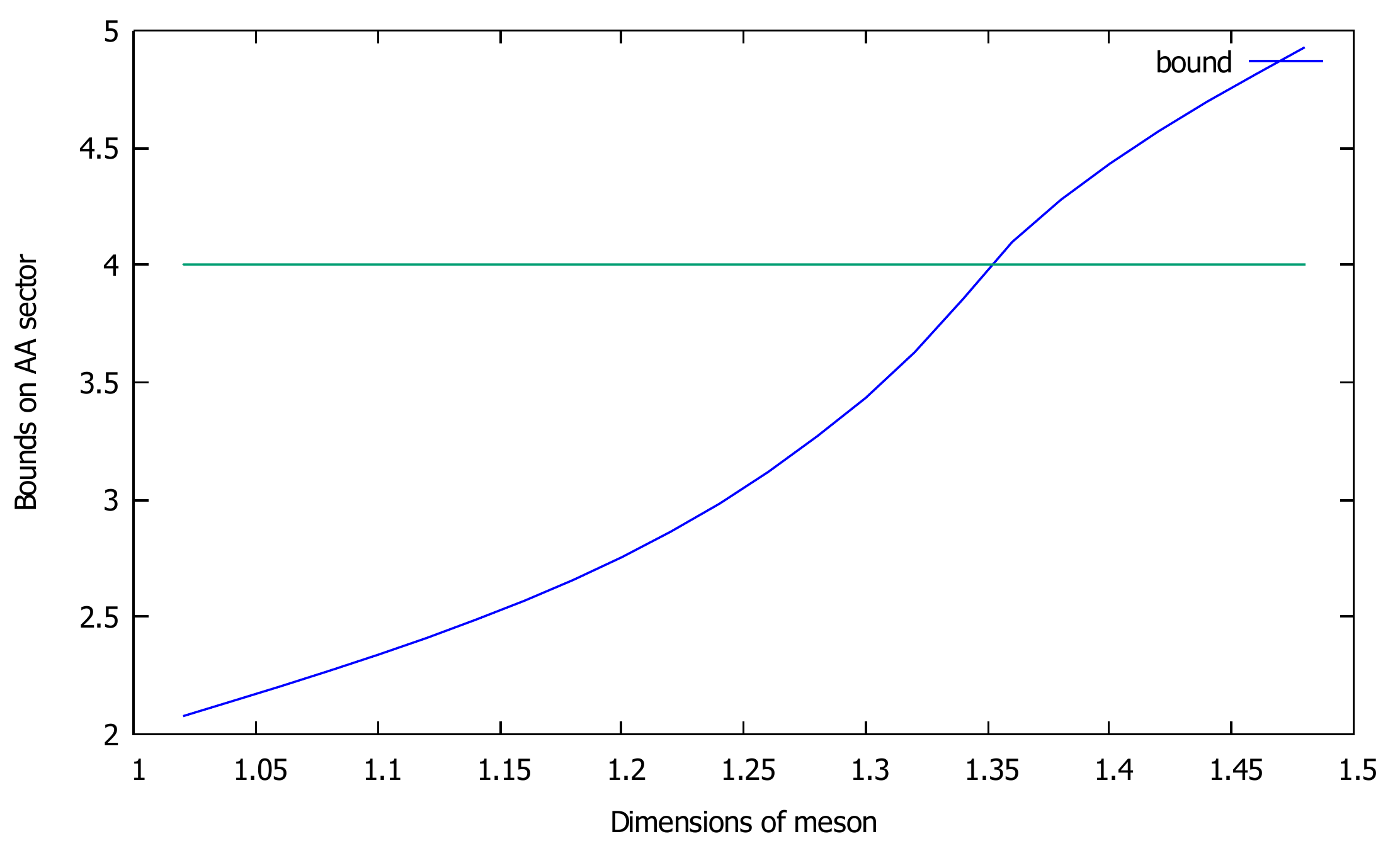}
  \end{center}
  \caption{Bounds on the scaling dimension of operators in the anti-symmetric $\times$ anti-symmetric representation.}
  \label{fig:aa}
\end{figure}

The resulting bound on the scaling dimensions can be found in Fig \ref{fig:tt} and Fig \ref{fig:aa}. Our result shows that in order to avoid these dangerously irrelevant operators, we need $\Delta_{\Phi} > 1.69$ or $\gamma_m < 1.31$ from the bound for the symmetric traceless $\times$ symmetric traceless representation in Fig \ref{fig:tt}, which is stronger than the one from the anti-symmetric $\times$ anti-symmetric representation.

\begin{figure}[htbp]
	\begin{center}
		\includegraphics[width=12.0cm,clip]{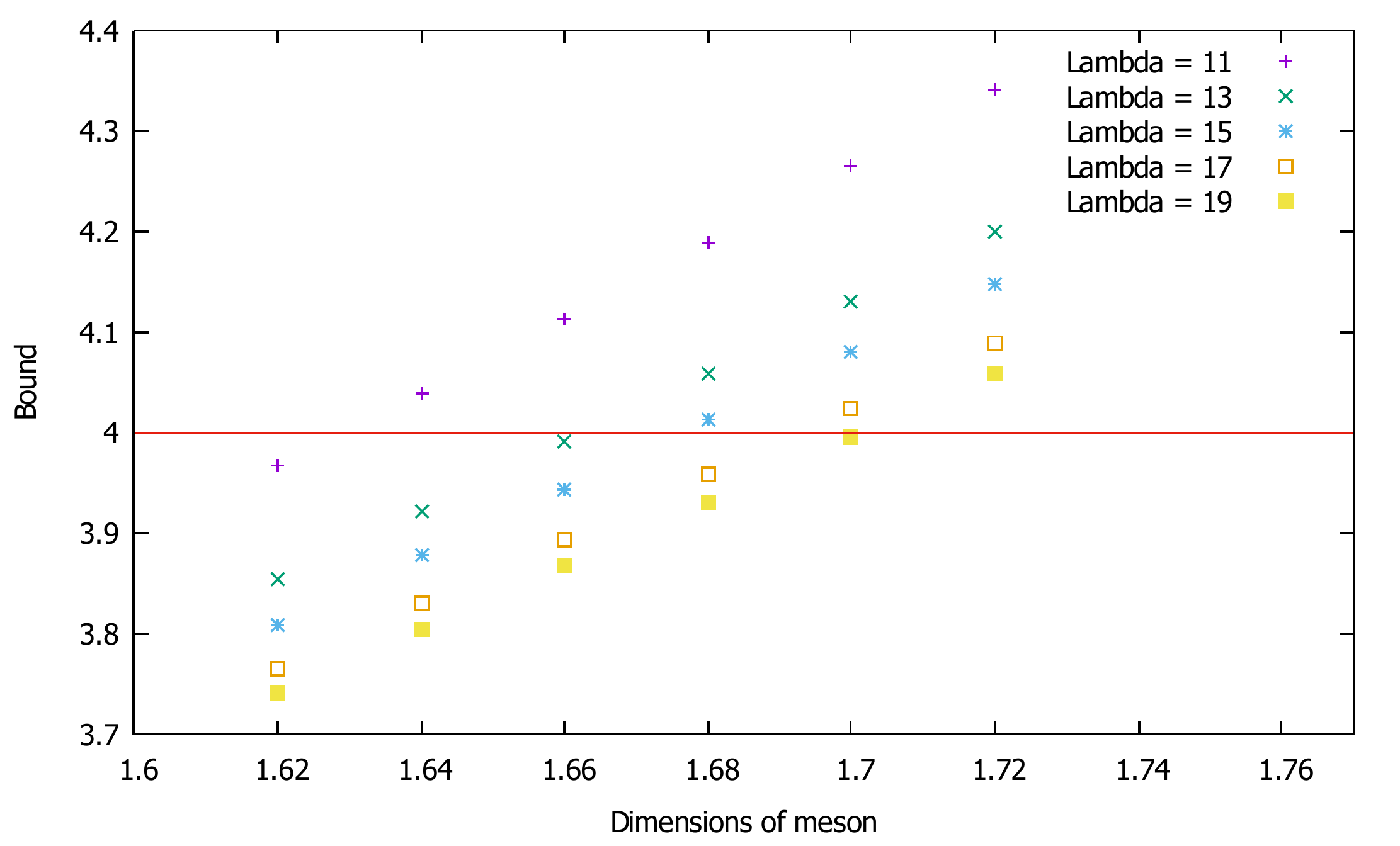}
	\end{center}
	\caption{Change of the bounds as we increase the search space dimension $\Lambda$}
	\label{fig:limiting}
\end{figure}

Our bound can be compared with the one studied in $SU(N_f)_V$ symmetric conformal field theories from the four-point functions of scalar operators in the adjoint representation \cite{Iha:2016ppj}. At first sight, it seems that our bound in Fig \ref{fig:tt} coincides with their bound in Fig A1, and our bound in Fig \ref{fig:aa} coincides with their bound in Fig A2. 
However, a more delicate comparison shows that they are very close but not identical if we use the same search space dimension $\Lambda$ for the comparison. It turns out that our bound on $\Delta_{\Phi}$ is slightly stronger (i.e. the bound on mass anomalous dimension $\gamma_m$ is stronger). As for the bound on the mass anomalous dimension for the possible phenomenological applications, however, the difference is minor in practice.\footnote{It is noted that the bound is very close to the mass anomalous dimension proposed in \cite{Ishikawa:2013tua}\cite{Ishikawa:2015iwa} from the Monte Carlo simulations based on Wilson fermions.} We also note that unlike in three dimensions near the kink, the convergence of the numerical conformal bootstrap bound with respect to the search space dimension $\Lambda$ is not dramatically fast in four dimensions, so one may expect a further improvement on the bound by increasing $\Lambda$. Indeed, we see certain degrees of improvement by increasing $\Lambda$ in Fig \ref{fig:limiting}, but we believe that the possible improvement on the bound must be much less than $0.1$ and does not reach $\gamma_m=1.0$. We will discuss the $\Lambda \to \infty$ limit further from a slightly different viewpoint in the next section.

\section{Discussions}
In this work, we have obtained an upper bound on the mass anomalous dimension in conformal multi-flavor QCD, assuming that the fixed point is realizable by regularization preserving the $SU(N_f)_V$ out of $SU(N_f)_L \times SU(N_f)_R$. The bound is numerically very similar but not identical to the one studied in \cite{Iha:2016ppj}. For $N_f=8$, our bound implies $\gamma_m < 1.31$ to avoid dangerously irrelevant operators that are not compatible with the lattice symmetry.

Is our bound saturated by the actual conformal multi-flavor QCD?
To address the point, we note that with the current technology of the conformal bootstrap, it is difficult to specify the gauge group, and our bound applies to any gauge theory with the same flavor symmetries. It is therefore very probable that our bound is not optimum, say for the study of the massless QCD with the $SU(3)$ gauge group. A possible idea to incorporate the gauge group is to study baryon operators or to study non-local observables, which may require further ingenuity from the viewpoint of the conformal bootstrap.\footnote{Recently, the similar question has been addressed in three-dimensional multi-flavor massless QED with the help of monopole operators \cite{Chester:2016wrc}\cite{Nakayama:2016jhq}.}

In relation to the above point, we have also studied the case with $N_f=16$. With the same search space dimensions (i.e. $\Lambda=17$) as in the case of $N_f=8$, we have found that the bound is  $\gamma_m < 1.29$. The choice of this $N_f$ is motivated by the Banks-Zaks fixed point: if we had studied the $SU(3)$ gauge group, one would find the very weakly coupled IR fixed point with the very small mass anomalous dimension $\gamma_m= 0.026$ computed in perturbation theory. It is obvious that the perturbative result satisfies our bound but does not saturate it. From this we should learn that our bounds are more useful in the strongly coupled regime in which the perturbative computation is not available.

To conclude the paper, we would like to discuss the large $N_f$ limit of our bound. In particular it is of academic interest if the bound becomes identical to the generalized free theory line in the large $N_f$ limit so that our bound on the mass anomalous dimension becomes  as small as $\gamma_m = 1$. This possibility is motivated from the observation that the bound in Fig \ref{fig:tt} may look a straight line for smaller $\Delta_{\Phi}$ and the slope becomes smaller for larger $N_f$. To motivate it further, we have empirically observed that the bound on the scaling dimension of the scalar operators in the symmetric traceless representation in $SO(N)$ symmetric conformal field theories in the large $N$ limit approaches the one for the generalized free theory line i.e. $\Delta_{T} = 2\Delta_{\phi}$ (at least in the range $\Delta_{T} < 4$). We do conjecture that the bound is actually the generalized free line in the large $N$ limit. 

\begin{figure}[htbp]
	\begin{center}
  \includegraphics[width=15.0cm,clip]{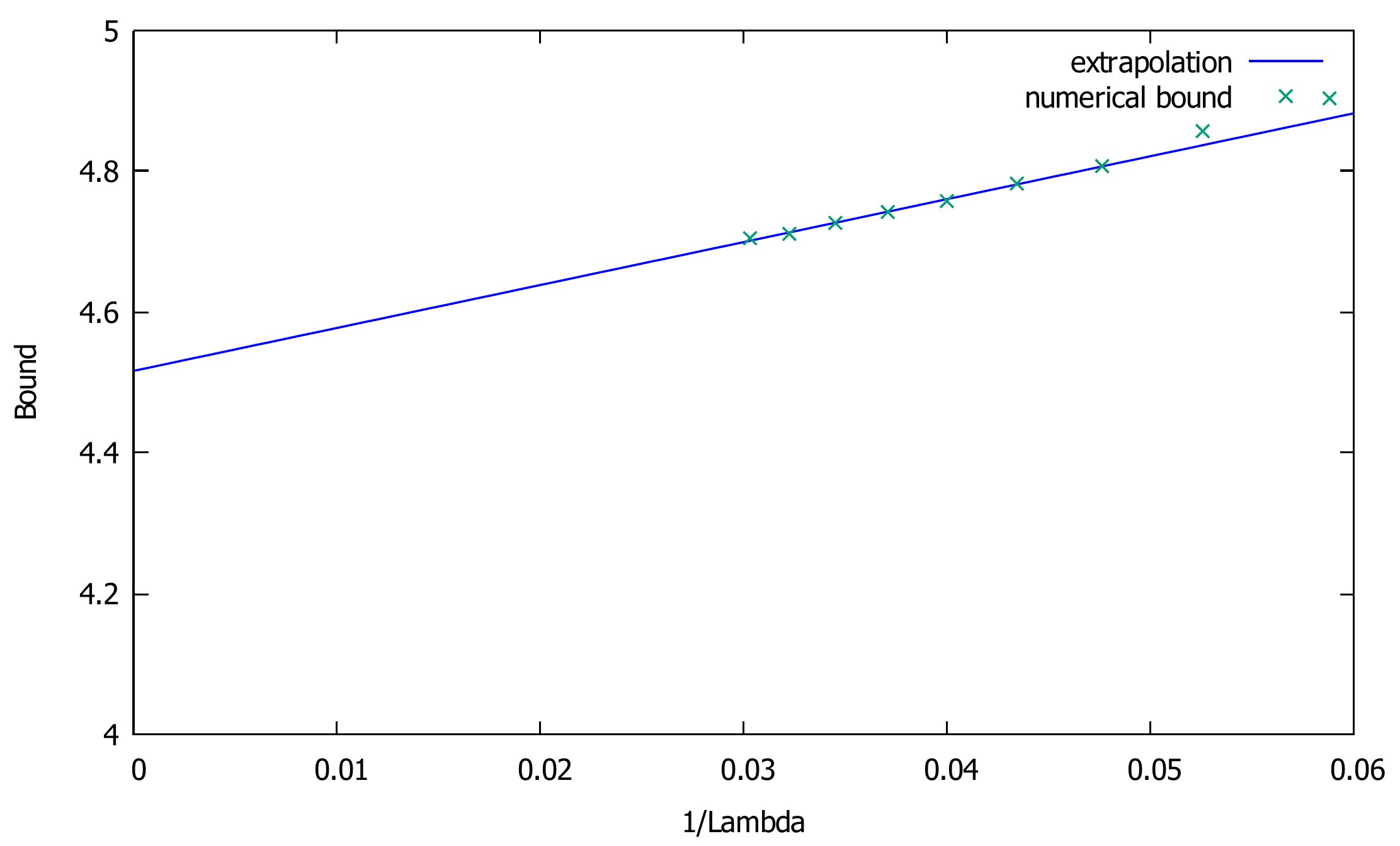}
  \end{center}
  \caption{The asymptotic behavior of the bound on the scaling dimensions of the scalar operators in the symmetric traceless $\times$ symmetric traceless representation in $SU(10000)_L \times SU(10000)_R$ symmetric conformal field theories as a function of $\Lambda^{-1}$.}
  \label{fig:limit}
\end{figure}

On the other hand, the bound on the scaling dimensions of the symmetric traceless $\times$ symmetric traceless  operator in $SU(N_f)_L \times SU(N_f)_R$ symmetric conformal field theories is weaker than the one of the  symmetric traceless representation in $SO(N)$ symmetric conformal field theories. While it may seem likely that for small $\Delta_{\Phi}$, the bound seems to approach the generalized free theory line in the large $N_f$ limit, we do conjecture that the bound on the scaling dimensions of scalar operators in the symmetric traceless $\times$ symmetric traceless representation in $SU(N_f)_L \times SU(N_f)_R$ symmetric conformal field theories is strictly weaker than the generalized free theory line.  To support our conjecture, in Fig \ref{fig:limit}, we present the bound at $\Delta_{\Phi} = 2.0$ with different values of $\Lambda$ with $N_f= 10000$. A crude extrapolation, in the spirit of \cite{Beem:2014zpa}\cite{Beem:2015aoa}, shows that the bound at $\Lambda = \infty$ tends to $\Delta_{TT} \sim 4.5$ which is above the generalized free theory one (i.e. 4.0). It should be contrasted with the case with $SO(N)$ theory bound in which $\Delta_{T}$ can be as small as $4.01$ or less even with $\Lambda \sim 20$ at $N=100000000$.

Finally, at the risk of repetition, we would like to stress that there is nothing wrong with conformal field theories, possibly including conformal multi-flavor QCD, possessing dangerously irrelevant operators, violating any of the bounds we have discussed. Indeed, we do know the existence of four-dimensional conformal field theories with $SU(N_f) \times SU(N_f)$ symmetry but violating our bounds with dangerously irrelevant operators. One example is the supersymmetric QCD in which the scaling dimension of chiral operators  in the fundamental $\times$ fundamental representation (i.e. supersymmetric meson operator) goes down to $\Delta_{\Phi} =1$ hitting the unitarity bound. Our bound does not contradict the case here, for the supersymmetric QCD does possess dangerously irrelevant operators. A part of the reason why such a fixed point may be stable is because supersymmetry protects the perturbation against it.

\section*{Acknowledgements}
The author would like to thank T.~Ohtsuki for valuable resources on numerical conformal bootstrap. Without his help, this research would not have been completed. He also thanks H.~Suzuki for discussions. 
A part of the numerical computation is done by a cluster server at Kavli IPMU. 


\end{document}